\documentclass[runningheads]{llncs}
\usepackage{cite}
\usepackage{amsmath,amssymb,amsfonts}
\usepackage{algorithmic}
\usepackage{graphicx}
\usepackage{textcomp}
\usepackage{subfigure}
\usepackage{hyperref}
\usepackage{microtype}

\usepackage{tikz}
\usetikzlibrary{positioning, arrows.meta}

\begin{document}

\title{Towards quantum machine learning for assessing the resilience of post-quantum cryptography}
\author{Jaroslaw A. Miszczak\orcidID{0000-0001-8790-101X}}	
\institute{Institute of Theoretical and Applied Informatics,
Polish Academy of Sciences,\\ Baltycka 5, 44-100 Gliwice, Poland\\
\email{jmiszczak@iitis.pl}}


\maketitle

\begin{abstract}
The potential capabilities of quantum computers motivated the development of cryptographic protocols suitable for securing communication against the adversaries with access to large fault-tolerant quantum computers. However, even though current quantum computers are limited in terms of size and precision, they can still be useful for finding loopholes and weaknesses in the post-quantum cryptographic protocols. In this work, we present an attempt  to utilized the capabilities of Quantum Generative Adversarial Networks (QGANs), one of the promising architectures used in quantum machine learning, for this purpose. We describe an example application of QGAN architecture for the purpose of loading the probability distribution of the hash-based digital signatures into the memory of a quantum computer. Our results confirm that near-term hybrid quantum-classical methods posses capabilities required for this purpose. The presented approach can be used as a first step in the workflow enabling the utilization of quantum computing for attacking post-quantum cryptographic primitives.

\end{abstract}

\section{Introduction}
\label{sec:introduction}
During the last decade, the anticipated capabilities of large-scale fault-tolerant quantum computers motivated the development of Post-Quantum Cryptography (PQC)~\cite{bernstein2017post, bernstein2009post}. PQC algorithms are specifically engineered to be secure against known quantum algorithms, primarily Shor's algorithm~\cite{shor1999polynomial} for public-key systems and Grover's algorithm~\cite{grover1996fast} for symmetric systems and hash functions. In particular, Shor's quantum algorithm for factorization demonstrated that quantum computers could be used to  efficiently solve mathematical problems like integer factorization and the computation of discrete logarithms, problems used as the basis of widely deployed public-key cryptosystems, including RSA~\cite{mosca2018cybersecurity}.

Such anticipated possibility gave rise to the \emph{harvest now -- decrypt later} scenario, also known as \emph{retrospective decryption}~\cite{nist-standard, bernstein2009post}, where adversaries may be intercepting and storing encrypted communications with the expectation of decrypting them in the future. However, for this threat to be realistic one has to  have for their disposal large-scale, fault-tolerant quantum computers, capable of running algorithms that can break widely used modern encryption methods. Such machines are usually described as Cryptographically Relevant Quantum Computers (CRQCs). The most important requirements for such machines are a large number of qubits, long coherence times, high-fidelity of quantum gates, and the ability to correct errors that occur during quantum computation.

In contrast, the current generation of quantum computing technology is represented by Noisy Intermediate-Scale Quantum (NISQ) computers. These devices are characterized by the limited number of qubits, usually of the order of hundreds, susceptibility to noise, resulting in the limited fidelity of gates, as well as short coherence times and limited ability to correct the errors. The most important class  of algorithms that can be run on NISQ devices and designed to be run on such devices are Variational Quantum Algorithms (VQAs)~\cite{cerezo2021variational}.

During the recent years, significant progress has been made in the adoption of the post-quantum cryptographic standards~\cite{nist-standard}, including major vendors of operating systems used in cloud environments~\cite{redhat-10}. Additionally, post-quantum algorithms have recently become available in OpenSSH 10.0 suite~\cite{openssh-10} which added support for a new hybrid post-quantum key exchange, based on the FIPS 203 standard. In March 2025, Java Platform version 24~\cite{java-24}, introduced post-quantum schemes for key encapsulation and digital signatures, making it one of the first general purpose, widely deployed programming languages adopting post-quantum cryptography. Java 26~\cite{java-26}, released in March 2026, introduces post-quantum-ready JAR signing and hybrid public key encryption support to prepare applications for the quantum era. 

The rapid progress in the adoption and standardization of post-quantum methods~\cite{chen2025standardisation} naturally makes them more available and, at the same time, more prone to developing new attacking techniques. In this regard, it is natural to ask if the parallel progress in the field of quantum technologies established a serious threat for the standardized post-quantum technologies. Indeed, only by investigating possible avenues for harnessing quantum computing for assessing the resilience of post-quantum protocols, it is possible to ensure their security.

In this work, we aim at tackling this problem by demonstrating that quantum computing techniques developed for near-term quantum computers can be used as a tool for post-quantum cryptography. In particular, we demonstrate that quantum machine learning can be utilized for as an initial step in the workflow where quantum computers are used to sample data encoded using post-quantum methods. We focus on a particular class of hybrid quantum-classical methods, namely Quantum Generative Adversarial Networks (QGANs). We also restrict our attention to a particular post-quantum protocol, namely the hash-based digital signature scheme.

The rest of this work is organized as follows. In~Section~\ref{sec:qgans-framework} we describe the basic elements used in Quantum Generative Adversarial Networks. In Section~\ref{sec:architecture} we introduce a hybrid architecture employed for learning the probability distribution, including data representation and quantum circuit topologies. Next, in Section~\ref{sec:results} we provide a result of the numerical experiments based on the post-quantum signature scheme and the described framework. Finally, in Section~\ref{sec:final} we summarize the contribution and provide some closing remarks.

\section{Quantum Generative Adversarial Networks}\label{sec:qgans-framework}

Classical Generative Adversarial Networks (GANs)~\cite{goodfellow2014generative} involve two neural networks -- a generator and a discriminator -- that are trained in an adversarial manner. The generator attempts to create synthetic data that mimics a target dataset, while the discriminator tries to distinguish between real data and the synthetic data produced by the generator.

Quantum Generative Adversarial Networks (QGANs)~\cite{dallairedemers2018quantum} represent a quantum computing extension of the classical GAN framework. In the quantum version of the generative adversarial network, the generator and the discriminator are replaced with quantum neural networks made up of parameterized quantum circuits. Another proposed architecture involves a quantum generator paired with a classical discriminator~\cite{huang2021power}.

The generator in QGAN takes a quantum state as input~\cite{havlicek2019supervised}, and applies a sequence of quantum gates to produce an output quantum state or classical data obtained through measurement. The major advantage of such design is that it enables the efficient encoding of complex probability distributions, which cannot be easily reproduced by the classical computer~\cite{arute2019quantum,huang2022quantum,benedetti2025complement}.

QGANs aim to leverage quantum phenomena such as superposition and entanglement to potentially enhance the generative capabilities and efficiency of GANs, opening up applications in quantum machine learning, quantum finance, and the generation of complex data distributions. It was demonstrated that the QGANs combining a variational quantum circuit and a classical neural network, can learn a representation of the probability distribution underlying the data samples and load it into a quantum state~\cite{zoufal2019quantum}. In \cite{romero2019variational} a hybrid quantum-classical approach to model continuous classical probability distributions using a variational quantum circuit is proposed. In \cite{benedetti2019adversarial} an adversarial algorithm for the problem of approximating an unknown quantum pure state is derived.

The preliminary proposals for employing classical and quantum machine learning (QML) for attacking post-quantum protocols can already be found in the literature. In \cite{dubrova2023breaking}, deep learning-based message recovery attacks on the $\omega$-order masked implementations of CRYSTALS-Kyber in ARM Cortex-M4 CPU are presented. In \cite{nguyenquantum}, the authors investigate adversarial risks in QML-assisted network functions and digital twin applications, including vulnerabilities such as quantum kernel poisoning, backdoor attacks, and adversarial noise.   In \cite{kim2023quantum}, an attack technique, based on the application of quantum machine learning to cryptoanalysis, for recovering keys in cryptographic algorithms is presented. In \cite{baksi2023new}, an extension of the ML-assisted differential attack model is presented, and it is argued that the traditional analysis of the differential distinguisher can lead to the underestimation of the true power of the ML-enabled attacker.  In \cite{s2024qhopnn}, the authors introduce the Quantum Hopfield Neural Network (QHopNN) as a novel approach to enhance key recovery in symmetric ciphers, and the proposed framework is evaluated using symmetric ciphers, including S-AES and S-DES, and benchmarked against the existing state-of-the-art techniques. In \cite{prasad2025safeguarding}, the authors highlight the intersection  of QGANs, PQC, and quantum key distribution (QKD), with possible  adversarial attack models on hybrid protocols. In  \cite{mohammad2025cyber}, GAN-like models are used to simulate quantum-resilient encryption and probe QKD flaws, and consider 
QGAN-inspired frameworks in network encryption security.

For the readers interested in the recent developments in the field of QGANs, we recommend \cite{tuan2023survey}, \cite{nokhwal2024quantum} and \cite{islam2025survey}. Some notable architectural variants are described in~\cite{zaman2023survey,pajuhanfard2024survey}. The method of using large language models (LLMs) for optimizing QGANs has been recently proposed in~\cite{ueda2025optimizing}.

\section{Framework for QGAN attacks on PQC}\label{sec:architecture}

Designing a concrete quantum circuit layout for a QGAN to test post-quantum cryptographic schemes involves a hybrid quantum-classical framework. In this section, we describe the data representation, the quantum circuit topology, parameter initialization, and the training loop.

\subsection{Architecture details}

\begin{figure*}
	\begin{center}
		\includegraphics[width=\textwidth]{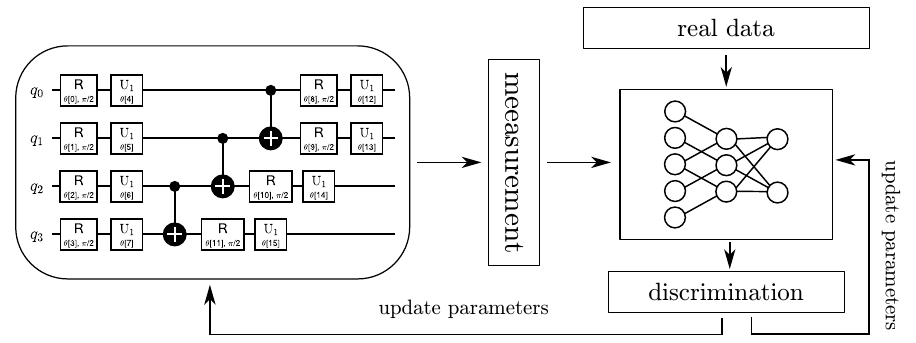}
	\end{center}
	\caption{Architecture of the proposed framework based on hybrid quantum-classical QGAN from the Qiskit Machine Learning library~\cite{sahin2025qiskit, qiskit-ml}. The quantum generator is trained using the feedback from the classical discriminator to generate a quantum state suitable for representing the observed real data. Details of the quantum circuit construction are described in Section~\ref{sec:quantum-architecture}. Details of the classical discriminator are described in Fig.~\ref{fig:discriminator-ann}.}
	\label{fig:architecture}
\end{figure*}

We will adapt the hybrid architecture which can be used to learn the 2D probability distribution described in~\cite{sahin2025qiskit,qiskit-ml}. The schematic description of the utilized architecture is presented in Fig.~\ref{fig:architecture}.

The quantum circuit handles the generator, while the discriminator can be either classical or quantum. The goal of the generator is to learn how to generate a probability distribution, obtained during the measurement process, in a way that is similar to the observed probability distribution from the real data.

Hence, the trained generator should be able to prepare an $n$-qubit pure quantum state of the form
\begin{equation}
	|\psi_{\text{t}}\rangle=\sum\limits_{j=0}^{k-1}\sqrt{p_{j}}|x_{j}\rangle,
\end{equation}
where the basis states $|x_{j}\rangle$ represent the data items in the training data set $X={x_0, \ldots, x_{k-1}}$ with $k\leq 2^n$ and $p_j$ is the probability of sampling element $x_{j}$. An element $x_{j}$ from the data set is represented by quantum state $|x_{j}\rangle$.

The role of the discriminator is to distinguish between the original distribution, observed in the real data, and the probabilities generated from the generator.
To train the generator and the discriminator we use the binary cross entropy as the loss function,
\begin{equation}
L\left(\boldsymbol{\theta}\right)=\sum_jp_j\left(\boldsymbol{\theta}\right)\left[y_j\log(x_j) + (1-y_j)\log(1-x_j)\right],
\end{equation}
where $x_j$ refers to a data sample and $y_j$ to the corresponding label. More details can be found in~\cite{qiskit-ml,sahin2025qiskit}.

As the result, the state prepared by the trained generator, represented by an ansatz and a vector of parameters, provides a quantum representation of the observed probability distribution. Hence, such a generator can be used to load the classical data into the quantum memory. 

\subsection{Data source and format}

For the purpose of this work, we analyse the data obtained as samples of the signatures  from SPHINCS+ hash-based digital signature scheme. As at the current stage the possibility of implementing QGANs -- on real quantum computers or in the simulators -- is limited, we analyse the probability distribution parts of the signatures only. We represent that data as pairs of four-bit words generated from the first byte of the signature. The example of the sample obtained from such a distribution is presented in Fig.~\ref{fig:observed-probs}.

\subsection{Quantum circuit topology}\label{sec:quantum-architecture}

\begin{figure}
	\begin{center}
		\subfigure[$N$-local ansatz with controlled-$R_y$, entanglement blocks and direct interaction between all qubits (full entanglement gates).]{
			\includegraphics[height=3cm]{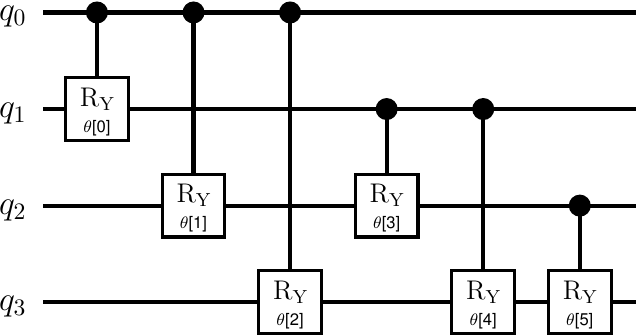}
			\label{fig:ansatz-n-local}
		}\quad %
			\subfigure[Real-amplitudes ansatz with linear entanglement gates.]{
			\includegraphics[height=3cm]{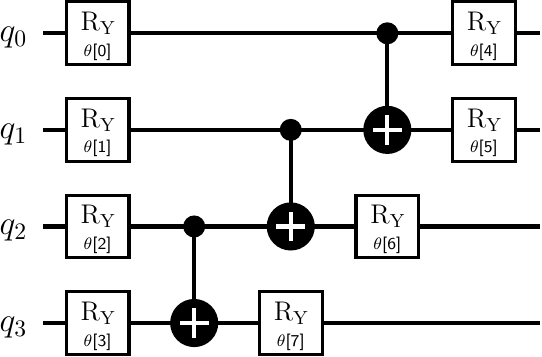}
			\label{fig:ansatz-real-amplitudes}
		}
	\end{center}	
	
	\begin{center}
		\subfigure[Hardware-efficient ansatz with linear entanglement gates.]{
			\includegraphics[height=3cm]{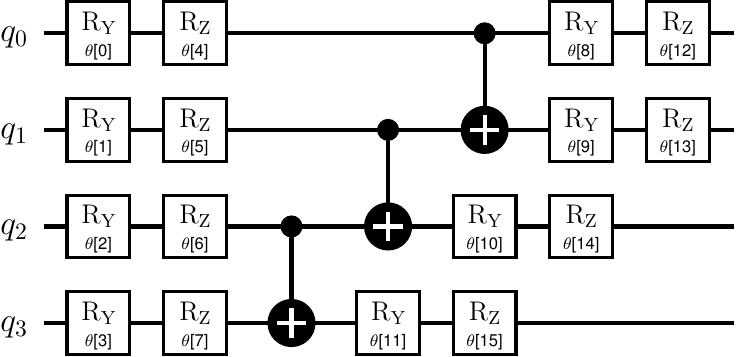}
			\label{fig:ansatz-efficient_su2}
		}
	\end{center}
	
	\caption{Three types of ansatze used in this work as the basic blocks of the quantum generator. The circuits in this figure were built for four-qubit registers. We have \subref{fig:ansatz-n-local} six, \subref{fig:ansatz-real-amplitudes} eight, and \subref{fig:ansatz-efficient_su2} sixteen trainable parameters per layer. 
		Using \subref{fig:ansatz-n-local} and \subref{fig:ansatz-efficient_su2} we do not impose restrictions on the form of generated quantum states.
		For the case \subref{fig:ansatz-real-amplitudes} the prepared quantum states will only have real amplitudes. }
	\label{fig:ansatz-examples}
\end{figure}

To implement the quantum generator, we use an ansatz provided by the Qiskit circuits library.  For the purpose of this study we use three types of quantum circuits, namely $N$-local ansatz, real-amplitudes ansatz, and hardware-efficient ansatz~\cite{qiskit-ml,sahin2025qiskit}. The examples of a single layer for each ansatz are presented in Fig.~\ref{fig:ansatz-examples}.

In particular, the structure of the $N$-local circuits is based on alternating rotation and entanglement layers.  Such circuits can often be simulated more efficiently on classical computers compared with other types of circuits. This is especially useful as it facilitates the development of new quantum algorithms, debugging, and verification before deployment on actual quantum hardware.

For real-amplitudes ansatz, the resulting circuit consists of alternating layers of $R_y$ and controlled negation (CNot) entangling gates. Such a pattern limits the expressibility~\cite{sim2019expressibility} of the circuit, understood as the ability to probe the space of quantum states, and only real amplitudes will be generated. 

In the case of hardware-efficient ansatz, the resulting circuit consists of layers of single-qubit operations spanned from $SU(2)$ and CNot gates. This ansatz is commonly used in variational quantum algorithms or classification circuits for machine learning. 

\subsection{Discriminator architecture}

\begin{figure}[ht!]
	\begin{center}
	\begin{tikzpicture}[
		node distance=.5cm,
		every node/.style={draw, rectangle, minimum height=1cm, minimum width=1.55cm, align=center},
		arrow/.style={->, thick}
		]
		
		\node (input) {Input \\ (2)};
		\node (linear1) [right=of input] {Linear \\ 2 $\to$ 20};
		\node (relu) [right=of linear1] {LeakyReLU \\ $\alpha=0.2$};
		\node (linear2) [right=of relu] {Linear \\ 20 $\to$ 1};
		\node (sigmoid) [right=of linear2] {Sigmoid};
		\node (output) [right=of sigmoid] {Output \\ (1)};
		
		\draw[arrow] (input) -- (linear1);
		\draw[arrow] (linear1) -- (relu);
		\draw[arrow] (relu) -- (linear2);
		\draw[arrow] (linear2) -- (sigmoid);
		\draw[arrow] (sigmoid) -- (output);
		
	\end{tikzpicture}

	\end{center}
	\caption{Architecture of the artificial neural network used as the discriminator in the presented method. The scheme is based on the implementation described in~\cite{sahin2025qiskit, qiskit-ml}.}
	\label{fig:discriminator-ann}
\end{figure}
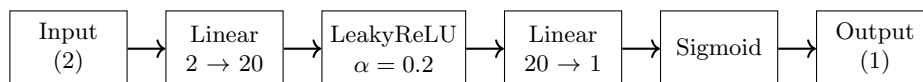

Discriminator in the presented method is implemented using PyTorch module for Python programming language. A detailed description of the architecture used by the discriminators is provided in Fig.~\ref{fig:discriminator-ann}.

\section{Results}\label{sec:results}

Let us now demonstrate the results of utilizing the defined framework for the purpose of loading the probability distribution of the signature bytes.

For the purpose of numerical experiments presented in this work, we used circuits built using six layers of the predefined ansatze, and the circuits were defined on a 16-qubit register. In this situation, we have 
56 trainable parameters for real-amplitudes ansatz, 98 parameters for $N$-local ansatz, and 112 parameters for hardware-efficient ansatz.

For the purpose of generating data used to test the presented workflow, we employ  PQCrypto~\cite{pqcrypto} module for Python programming language. This module provides bindings to implementations of quantum-resistant cryptographic algorithms that were submitted to the NIST Post-Quantum Cryptography Standardization process. PQCrypto is based on C implementations derived from the PQClean project~\cite{kannwischer2022improving}.

\begin{figure}[ht!]
	\begin{center}
		\subfigure[Real data from SPHINCS+ signatures.]{
			\includegraphics[width=0.45\columnwidth]{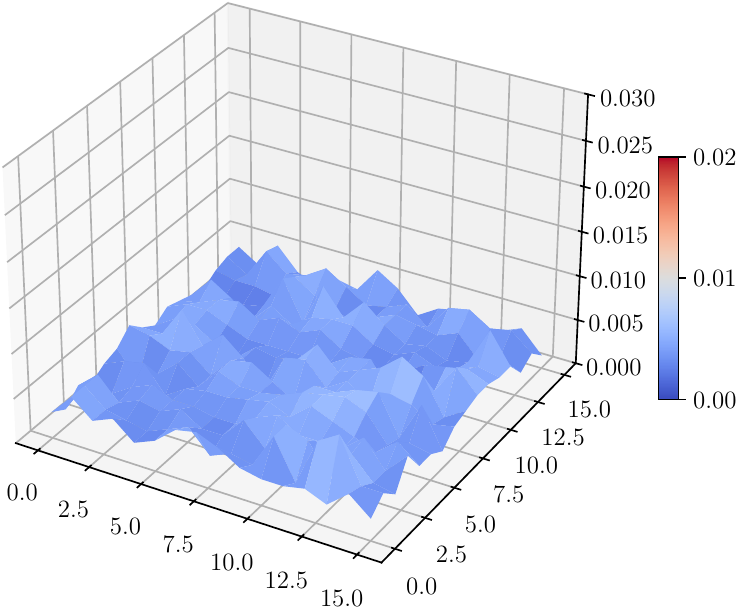}
			\label{fig:observed-probs}
		}
\hspace{0.025\columnwidth}
		\subfigure[Mixture of multivariate normal distributions.]{
			\includegraphics[width=0.45\columnwidth]{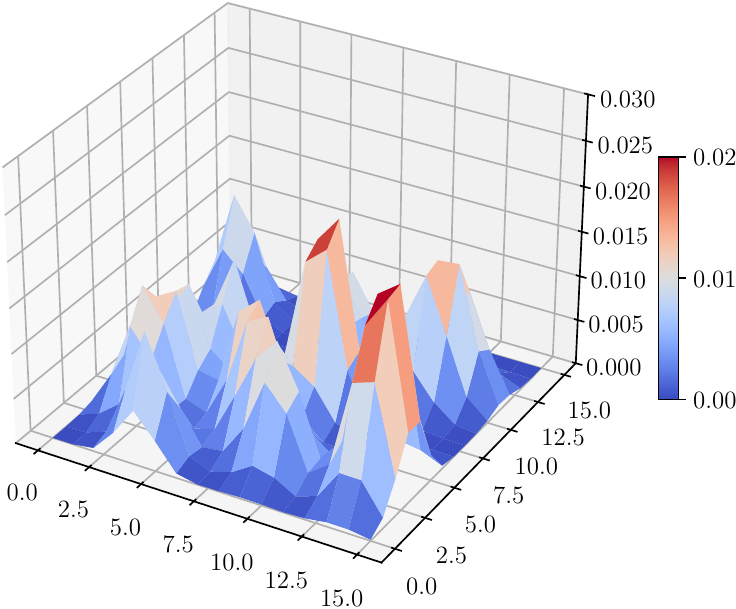}
			\label{fig:multivariate-normal-mix}
			}
	\end{center}
\caption{Sample data from  \subref{fig:observed-probs} the post-quantum  hash-based source and \subref{fig:multivariate-normal-mix} an artificially created mixture of  multivariate normal distributions. \subref{fig:observed-probs} Probability distribution of pairs of four-bit words generated from the first byte of the SPHINCS+ signatures. A sample obtained using $2^{13}$ signatures for a single plain text. \subref{fig:multivariate-normal-mix} Probability distribution obtained as a mixture of 16 multivariate normal distributions, with parameters adjusted to  $(0,15)\times (0,15)$ grid.}
\end{figure}

Parameter optimization loop was based on ADAM optimizer~\cite{kinga2015method}.
The optimization of the circuit parameters is achieved using the optimizer with learning rate $0.01$ and parameters $\beta_1=0.7$, $\beta_2=0.999$.

\begin{figure}[h!]
	\begin{center}
		\includegraphics[width=\columnwidth]{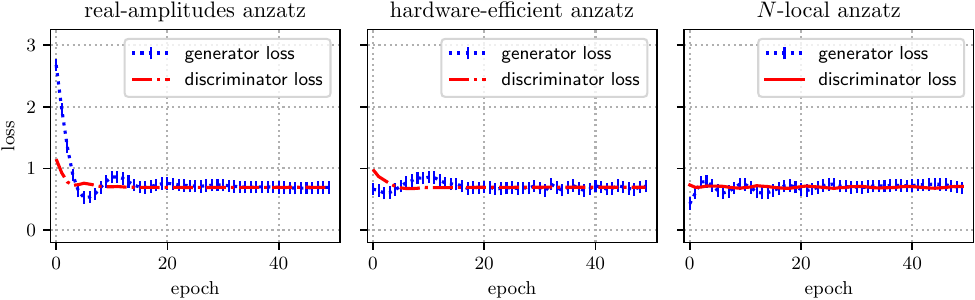}	
	\end{center}
	\caption{Value of the loss function for the generator and the discriminator during the learning process for the PQC data.}
	\label{fig:loss-values}
\end{figure}

The values of the loss function for three types of quantum circuits are shown in Fig~\ref{fig:loss-values}. The value of the loss function indicates convergence, signalling that the model's parameters have settled into a stable state.

For the purpose of quantifying the quality of the obtained approximation we use the Kullback-Leibler divergence, which provides a measure of how much the probability distributions differ. This quantity is defined as the relative entropy between the probability distributions
\begin{equation}
	D_\text{KL}(P \parallel Q) = \sum_{ x \in \mathcal{X} } P(x) \, \log \frac{ P(x) }{ Q(x) }.
\end{equation}

\begin{figure}[ht!]
	\begin{center}
	\includegraphics[scale=0.9]{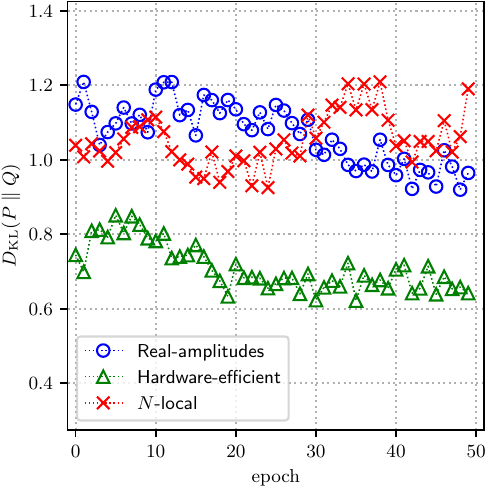}	
\end{center}
	\caption{Distance between the learned probability distribution and the observed probability distribution for the data from SPHINCS+, a post-quantum secure, hash-based digital signature scheme from the NIST PQC suite.}
	\label{fig:entropy-values}
\end{figure}

\begin{figure}[ht!]
\begin{center}
	\includegraphics[scale=0.9]{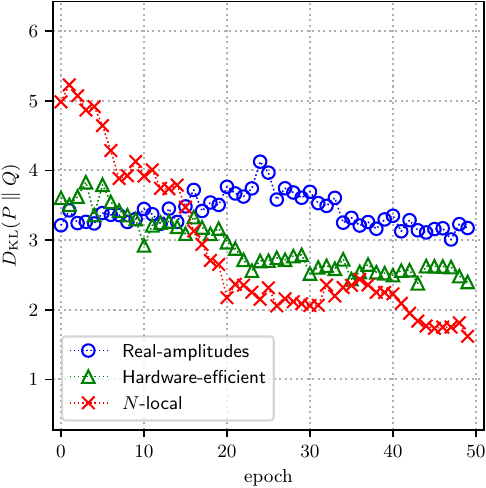}
\end{center}	
	\caption{Distance between the learned probability distribution and the observed probability distribution for the data from the distribution obtained as a mixture of 16 multivariate normal distributions.}
	\label{fig:normal-entropy-values}
\end{figure}

The results of evaluating Kullback-Leibler divergence for the quantum generators constructed using different ansatze are presented in~Fig.~\ref{fig:entropy-values}. From the obtained results one can conclude that the proposed method is able to learn the probability distribution of the signatures in the case of hardware-efficient ansatz and real-amplitudes ansatz. For the sake of completeness, in Fig.~\ref{fig:normal-entropy-values} we plot the values of Kullback-Leibler divergence for the mixture of multivariate normal distributions.

Additionally, one can note an important point. The results of hardware-efficient ansatz are better and lead to a more accurate approximation of the real distribution. This can be partially explained by the larger number of trainable parameters (112 in our case) compared to two other cases. However, one should also note that real-amplitudes ansatz, despite using only 56 trainable parameters, was able to deliver a better approximation of the real data compared with $N$-local ansatz (using 96 trainable parameters). Hence, the ability of the ansatz to probe the space of quantum states is not always beneficial for the task of learning real-world data.

To benchmark the effectiveness of the proposed method in the case of the less complex probability distribution, we use the artificially created mixture of multivariate normal distributions. An example of a sample from this distribution is presented in Fig.~\ref{fig:multivariate-normal-mix}. The task of learning this distribution is based on the task from~\cite{qiskit-ml}. Still, one should note that this task is presented for the purpose of illustrating the  differences in the effectiveness of the used ansatze.

From the results  in Fig.~\ref{fig:multivariate-normal-mix} one can note that in the case of the mixture of normal distributions, all architectures used to construct a quantum generator are equally effective during the process of distribution learning. This is equivalent to saying that -- from the point of view of quantum circuit construction -- this task is easier than the task of learning PQC data. This is especially visible as, in this case, the circuit based on hardware-efficient ansatz does not have any advantages over other ansatze. However, the least powerful ansatz, namely the real-amplitude ansatz, leads to the lower quality of approximation. Hence, the advantage in the circuit expressibility can lead to better results, but it is not directly connected with the learning capabilities of the QGAN.

As the last observation, one can note that for both learning the PQC data and the normal distribution, the real-amplitudes ansatz provides a solid method for learning the distribution. As, in this case, the quantum states are limited to real-amplitudes only, one can conclude that such restriction does not limit the power of the quantum generator. Hence, even if this ansatz does not prove to be the best candidate from the perspective of the approximation quality, it can be used as a less computationally demanding alternative for simple tasks.

\section{Conclusions}\label{sec:final}
The goal of this work was to provide an overview of the quantum machine learning approach for supporting the assessment of the security of post-quantum cryptographic protocols. The presented results confirm that Quantum Generative  Adversarial Networks posses generalization capabilities required to reproduce complex probability distribution generated by the post-quantum cryptography schemes. Although the current feasibility of attacks harnessing NISQ quantum computers remains limited, one should consider the growing power of such machines, and subsequently the growing possibilities of quantum variational algorithms, as a serious threat.

The potential for applying QGANs in cryptoanalysis stems from their ability to learn complex data distributions and generate synthetic data that closely resembles the training data. In the context of attacking PQC protocols, this capability could theoretically be leveraged in several ways~\cite{nokhwal2024quantum}.

If a QGAN could be trained to accurately model the distribution from which secret keys are drawn, it might then be able to generate candidate keys that have a higher likelihood of being correct. This could effectively reduce the search space for brute-force attacks, potentially making them feasible even against the protocols designed to resist such attacks classically.

Furthermore, QGANs, similar to their classical counterparts, might be employed to identify subtle vulnerabilities or non-random patterns in the outputs or internal states of PQC algorithms. Just as classical GANs have been used to find weaknesses in other neural networks, QGANs could potentially uncover unforeseen flaws in the design or implementation of PQC protocols, leading to the development of more efficient attacks that exploit these weaknesses.

The main advantage of the procedure described in this work is that it requires $\mathcal{O}\left(poly\left(n\right)\right)$ gates. Hence, it enables the loading of the probability distribution underlying the signatures into a quantum channel.  This, in turn, enables us to utilize potentially advantageous quantum algorithms -- including the variational quantum algorithms -- for processing the classical data. Thanks to this, the presented method, due to the ability of quantum computers to generate probability distribution with complex correlation, has the potential to uncover non-obvious leakages and recognize the weaknesses of post-quantum algorithms.

However, one should note that the framework presented in this work is limited by the current state of the quantum computing technology. In particular, QGANs are sensitive to noise and the current quantum computers are still lacking the level of error-correction required to implement them with high fidelity. Furthermore, as one can note from the different quality of the results obtained using different ansatze, training QGANs is a non-trivial task as it requires a precise and time-consuming simulations. 

One should also note that the presented experiments are based on the simulation and the considered dataset consists only of a small portion of signature data. This is mostly motivated by the limited access for quantum computers with suitable computing power. Hence the presented study does not present a full cryptanalytic attack, such as key recovery or message forgery, thus limiting its immediate practical implications.

Nevertheless, as quantum computers with a larger number of qubits, longer coherence times, and lower error rates become a reality, the ability to implement more complex quantum algorithms, including sophisticated QGANs, will increase.  Future attack strategies might also involve sophisticated hybrid quantum-classical approaches. In such scenarios, QGANs could potentially be employed as a component within a larger attack framework that strategically combines the strengths of both classical and quantum computational resources. For this reason, it is crucial to explore the potential for both offensive and defensive applications of quantum machine learning in the context of ensuring secure communication in the face of evolving quantum computational capabilities. This work provides a first step in this direction.

\section*{Acknowledgments}

This work was partially supported by EU grant Q-FENCE "Securing Tomorrow’s Digital Infrastructure with Quantum-Resistant Cryptography" (project number 101225708). Source code used for the numerical experiments presented in this work is based on the implementation of QGAN described in~\cite{sahin2025qiskit}, available from~\cite{qiskit-ml}. Author would like to thank Izabela Miszczak for proofreading the manuscript and to anonymous reviews for providing constrictive remarks.


\begin{thebibliography}{10}
	\providecommand{\url}[1]{\texttt{#1}}
	\providecommand{\urlprefix}{URL }
	\providecommand{\doi}[1]{https://doi.org/#1}
	
	\bibitem{openssh-10}
	{OpenSSH 10. Release Notes} (Apr 2025),
	\url{https://www.openssh.com/txt/release-10.0}
	
	\bibitem{nist-standard}
	Post-quantum cryptography standardization (May 2025),
	\url{https://csrc.nist.gov/pqc-standardization}
	
	\bibitem{pqcrypto}
	pqcrypto (v.0.3.4): {Post-quantum cryptography for Python} (Jul 2025),
	\url{https://pypi.org/project/pqcrypto/}
	
	\bibitem{qiskit-ml}
	Qiskit machine learning (2025),
	\url{https://github.com/qiskit-community/qiskit-machine-learning}
	
	\bibitem{java-24}
	{Reference Implementation of version 24 of the Java SE Platform} (Mar 2025),
	\url{https://openjdk.org/projects/jdk/24/}
	
	\bibitem{java-26}
	{Reference Implementation of version 26 of the Java SE Platform} (Mar 2026),
	\url{https://openjdk.org/projects/jdk/26/}
	
	\bibitem{arute2019quantum}
	Arute, F., et~al.: Quantum supremacy using a programmable superconducting
	processor. Nature  \textbf{574}(7779),  505--510 (Oct 2019).
	\doi{10.1038/s41586-019-1666-5}
	
	\bibitem{baksi2023new}
	Baksi, A., Breier, J., Dasu, V.A., Hou, X., Kim, H., Seo, H.: New results on
	machine learning-based distinguishers. IEEE Access  \textbf{11},
	54175--54187 (2023). \doi{10.1109/access.2023.3270396}
	
	\bibitem{benedetti2025complement}
	Benedetti, M., Buhrman, H., Weggemans, J.: Complement sampling: Provable,
	verifiable and nisqable quantum advantage in sample complexity (2025).
	\doi{10.48550/ARXIV.2502.08721}
	
	\bibitem{benedetti2019adversarial}
	Benedetti, M., Grant, E., Wossnig, L., Severini, S.: Adversarial quantum
	circuit learning for pure state approximation. New Journal of Physics
	\textbf{21}(4),  043023 (Apr 2019). \doi{10.1088/1367-2630/ab14b5}
	
	\bibitem{bernstein2009post}
	Bernstein, D.J., Buchmann, J., Dahmen, E. (eds.): Post-Quantum Cryptography.
	Springer Berlin Heidelberg, Berlin, Heidelberg (2009)
	
	\bibitem{bernstein2017post}
	Bernstein, D.J., Lange, T.: Post-quantum cryptography. Nature
	\textbf{549}(7671),  188--194 (Sep 2017). \doi{10.1038/nature23461}
	
	\bibitem{cerezo2021variational}
	Cerezo, M., Arrasmith, A., Babbush, R., Benjamin, S.C., Endo, S., Fujii, K.,
	McClean, J.R., Mitarai, K., Yuan, X., Cincio, L., Coles, P.J.: Variational
	quantum algorithms. Nature Reviews Physics  \textbf{3}(9),  625--644 (Aug
	2021). \doi{10.1038/s42254-021-00348-9}
	
	\bibitem{chen2025standardisation}
	Chen, L.: Standardisation of and Migration to Post-Quantum Cryptography,
	vol. International Conference on Research in Security Standardisation, pp.
	3--13. Springer Nature (2025). \doi{10.1007/978-3-031-87541-0_1}
	
	\bibitem{dallairedemers2018quantum}
	Dallaire-Demers, P.L., Killoran, N.: Quantum generative adversarial networks.
	Physical Review A  \textbf{98}(1),  012324 (Jul 2018).
	\doi{10.1103/physreva.98.012324}
	
	\bibitem{dubrova2023breaking}
	Dubrova, E., Ngo, K., Gärtner, J., Wang, R.: Breaking a fifth-order masked
	implementation of crystals-kyber by copy-paste. In: Proceedings of the 10th
	ACM Asia Public-Key Cryptography Workshop. pp. 10--20. ASIA CCS ’23, ACM
	(Jul 2023). \doi{10.1145/3591866.3593072}
	
	\bibitem{goodfellow2014generative}
	Goodfellow, I.J., Pouget-Abadie, J., Mirza, M., Xu, B., Warde-Farley, D.,
	Ozair, S., Courville, A., Bengio, Y.: Generative adversarial networks (2014).
	\doi{10.48550/arXiv.1406.2661}
	
	\bibitem{grover1996fast}
	Grover, L.K.: A fast quantum mechanical algorithm for database search. In:
	Proc. 28th annual ACM symposium on Theory of computing. pp. 212--219 (1996).
	\doi{10.1145/237814.237866}
	
	\bibitem{s2024qhopnn}
	Hariharasitaraman, S., Mishra, N., Vishnuvardhanan, D.: {QHopNN}: investigating
	quantum advantage in cryptanalysis using a quantum hopfield neural network.
	Physica Scripta  \textbf{99}(8),  086002 (Jul 2024).
	\doi{10.1088/1402-4896/ad5ed1}
	
	\bibitem{havlicek2019supervised}
	Havlíček, V., Córcoles, A.D., Temme, K., Harrow, A.W., Kandala, A., Chow,
	J.M., Gambetta, J.M.: Supervised learning with quantum-enhanced feature
	spaces. Nature  \textbf{567}(7747),  209--212 (Mar 2019).
	\doi{10.1038/s41586-019-0980-2}
	
	\bibitem{huang2022quantum}
	Huang, H.Y., Broughton, M., Cotler, J., Chen, S., Li, J., Mohseni, M., Neven,
	H., Babbush, R., Kueng, R., Preskill, J., McClean, J.R.: Quantum advantage in
	learning from experiments. Science  \textbf{376}(6598),  1182--1186 (Jun
	2022). \doi{10.1126/science.abn7293}
	
	\bibitem{huang2021power}
	Huang, H.Y., Broughton, M., Mohseni, M., Babbush, R., Boixo, S., Neven, H.,
	McClean, J.R.: Power of data in quantum machine learning. Nature
	Communications  \textbf{12}(1) (May 2021). \doi{10.1038/s41467-021-22539-9}
	
	\bibitem{islam2025survey}
	Islam, M., Turkeli, S., Ozaydin, F.: A survey of quantum generative adversarial
	networks: Architectures, use cases, and real-world implementations  (2025).
	\doi{10.48550/arXiv.2506.18002}
	
	\bibitem{kannwischer2022improving}
	Kannwischer, M.J., Schwabe, P., Stebila, D., Wiggers, T.: Improving software
	quality in cryptography standardization projects. In: {IEEE} European
	Symposium on Security and Privacy, EuroS{\&}P 2022 - Workshops, Genoa, Italy,
	June 6-10, 2022. pp. 19--30. IEEE Computer Society, Los Alamitos, CA, USA
	(2022). \doi{10.1109/EuroSPW55150.2022.00010}
	
	\bibitem{kim2023quantum}
	Kim, H., Lim, S., Baksi, A., Kim, D., Yoon, S., Jang, K., Seo, H.: Quantum
	artificial intelligence on cryptanalysis. Cryptology ePrint Archive  (2023),
	\url{https://ia.cr/2023/004}
	
	\bibitem{kinga2015method}
	Kinga, D., Adam, J.B., et~al.: A method for stochastic optimization. In:
	International conference on learning representations (ICLR). vol.~5. San
	Diego, California (2015), \url{https://arxiv.org/abs/1412.6980}
	
	\bibitem{mohammad2025cyber}
	Mohammad, K.: Cyber Shield: Advances in Detection, Isolation, and Containment
	Mechanisms. American Institute of Aeronautics and Astronautics (Jan 2025).
	\doi{10.2514/6.2025-2724},
	\url{https://arc.aiaa.org/doi/abs/10.2514/6.2025-2724}
	
	\bibitem{mosca2018cybersecurity}
	Mosca, M.: Cybersecurity in an era with quantum computers: Will we be ready?
	IEEE Security \& Privacy  \textbf{16}(5),  38--41 (Sep 2018).
	\doi{10.1109/msp.2018.3761723}, \url{https://ia.cr/2015/1075}
	
	\bibitem{tuan2023survey}
	Ngo, T.A., Nguyen, T., Thang, T.C.: A survey of recent advances in quantum
	generative adversarial networks. Electronics  \textbf{12}(4) (2023).
	\doi{10.3390/electronics12040856}
	
	\bibitem{nguyenquantum}
	Nguyen, V.L., Nguyen, L.H., Hwang, R.H., Canberk, B., Duong, T.Q.: Quantum
	machine learning for {6G} network intelligence and adversarial threats. IEEE
	Communications Standards Magazine  \textbf{9} (Sep 2025).
	\doi{10.1109/MCOMSTD.2025.3575261}
	
	\bibitem{nokhwal2024quantum}
	Nokhwal, S., Nokhwal, S., Pahune, S., Chaudhary, A.: Quantum generative
	adversarial networks: Bridging classical and quantum realms. In: 2024 8th
	International Conference on Intelligent Systems, Metaheuristics \& Swarm
	Intelligence (ISMSI). pp. 105--109. ISMSI 2024, ACM (Apr 2024).
	\doi{10.1145/3665065.3665082}
	
	\bibitem{pajuhanfard2024survey}
	Pajuhanfard, M., Kiani, R., Sheng, V.S.: Survey of quantum generative
	adversarial networks (qgan) to generate images. Mathematics  \textbf{12}(23),
	~3852 (2024). \doi{10.3390/math12233852}
	
	\bibitem{prasad2025safeguarding}
	Prasad, R., Koren, A.: Safeguarding {6G}: Security and Privacy for the Next
	Generation. River Publishers (May 2025)
	
	\bibitem{redhat-10}
	{Red Hat Inc.}: {4 key steps to prepare for post-quantum cryptography} (May
	2025),
	\url{https://www.redhat.com/en/resources/4-steps-for-postquantum-cryptography-checklist}
	
	\bibitem{romero2019variational}
	Romero, J., Aspuru-Guzik, A.: Variational quantum generators: Generative
	adversarial quantum machine learning for continuous distributions (2019).
	\doi{10.48550/arXiv.1901.00848}
	
	\bibitem{sahin2025qiskit}
	Sahin, M.E., Altamura, E., Wallis, O., Wood, S.P., Dekusar, A., Millar, D.A.,
	Imamichi, T., Matsuo, A., Mensa, S.: Qiskit machine learning: an open-source
	library for quantum machine learning tasks at scale on quantum hardware and
	classical simulators  (May 2025), \url{https://arXiv.org/abs/2505.17756}
	
	\bibitem{shor1999polynomial}
	Shor, P.W.: Polynomial-time algorithms for prime factorization and discrete
	logarithms on a quantum computer. SIAM Review  \textbf{41}(2),  303--332 (Jan
	1999). \doi{10.1137/s0036144598347011}
	
	\bibitem{sim2019expressibility}
	Sim, S., Johnson, P.D., Aspuru-Guzik, A.: Expressibility and entangling
	capability of parameterized quantum circuits for hybrid quantum-classical
	algorithms. Advanced Quantum Technologies  \textbf{2}(12),  1900070 (2019)
	
	\bibitem{ueda2025optimizing}
	Ueda, K., Matsuo, A.: Optimizing ansatz design in quantum generative
	adversarial networks using large language models (2025).
	\doi{10.48550/arXiv.2503.12884}
	
	\bibitem{zaman2023survey}
	Zaman, K., Marchisio, A., Hanif, M.A., Shafique, M.: A survey on quantum
	machine learning: Current trends, challenges, opportunities, and the road
	ahead (2023). \doi{10.48550/arXiv.2310.10315}
	
	\bibitem{zoufal2019quantum}
	Zoufal, C., Lucchi, A., Woerner, S.: Quantum generative adversarial networks
	for learning and loading random distributions. npj Quantum Information
	\textbf{5}(1) (Nov 2019). \doi{10.1038/s41534-019-0223-2}
	
\end{thebibliography}

\end{document}